# Selective surface modification and layer thinning of $MoS_2$ via ultraviolet light irradiation in ionic solution

*Lei Zhang,[†] Kun Liu,[†] Fanyi Kong,[†] Xue Han,[†] Jianxun Dai,[†] Mengmeng Wang,[#],\* Changsen Sun,[†] Huolin Huang,[†] Lujun Pan,[§] and Dawei Li[†],\**

[†] School of Optoelectronic Engineering and Instrumentation Science, Dalian University of Technology, Dalian 116024, China

[#] Laser Micro/Nano Fabrication Laboratory, School of Mechanical Engineering, Beijing Institute of Technology, Beijing 100081, China

[§] School of Physics, Dalian University of Technology, Dalian 116024, China

\* Correspondence to: mwang6@bit.edu.cn; dwli@dlut.edu.cn

**KEYWORDS**: *transition-metal dichalcogenide, surface modification, p-type doping, layer thinning, photo-reduction*

**ABSTRACT**

The electrical and optoelectronic properties of transition-metal dichalcogenides (TMDs), such as $MoS_2$, are highly dependent on carrier doping and layer thickness. The ability to selectively control these two critical characteristics is of great importance to develop TMD-based multifunctional device applications, which remains challenging. Here, we report a strategy for controllable surface modification and layer thinning of $MoS_2$ via ultraviolet (UV) light irradiation in a silver ionic solution environment. The results show that by adjusting UV irradiation time, nanostructured silver ultrathin films (~2.9 nm) are uniformly deposited on monolayer $MoS_2$ and can lead to controllable *p*-type doping effect, while the thickness of $MoS_2$ from few-layer to bulk crystals could be thinned down to the atomic monolayer limit. Both silver nanostructure deposition and layer thinning process have been evidenced to initiate from the edges of $MoS_2$, and independent of the edge type, thus revealing a unique UV light-assisted defect-induced surface modification and layer thinning mechanism. Overall, this study provides a new methodology for selective control of doping and layer thickness in TMDs, paving the way for developing novel 2D nanoelectronics and integrated optoelectronics.

**INTRODUCTION**

Owning to their excellent electrical and optoelectronic properties, two-dimensional (2D) semiconducting transition-metal dichalcogenides (TMDs), such as $MoS_2$, show great potential in diverse device applications, including field-effect transistors (FETs),[1-3] photodetectors,[4] nanocavity lasers,[5] optical modulators,[6-7] sensors,[8] and energy storage devices.[9] To realize these applications, it is necessary to control their electrical and optical properties, both of which are highly dependent on the number of layers[10-11] and the surface doping.[12-13] For instance, TMDs from few-layer to bulk materials have indirect band gap with weak luminescence, while monolayer (1L) TMDs exhibit direct bandgap with strong luminescence,[14-15] and also possess other unique properties such as large second-order optical nonlinearities,[16] strong spin-valley coupling,[17] etc. Besides, the doping as a simple and most commonly functionality way, can not only adjust the band gap but also modulate the conductivity in layered TMDs.[18]

In the previous studies, the number of layers (especially the monolayer limit) in TMDs has been controlled via mechanical exfoliation,[19] chemical vapor deposition,[20] plasma etching,[21] and laser thinning.[22-23] On the other hand, several doping techniques have been developed to modulate the inherent properties of TMDs, including chemical doping,[24] single-atom metal doping,[25] laser-assisted doping,[26] and surface modification.[13] However, most of these techniques are complex and limited to only one characteristic control. The ability to control the layer thickness and the doping through a single-platform approach is of great importance to develop TMD-based multifunctional and integrated device applications, which remains challenging and yet to be demonstrated.

A promising approach to achieve both carrier doping and layer thickness control is light-assisted post-treatment of layered TMDs.[27-28] In this work, we exploit ultraviolet (UV) light irradiation to selectively realize surface modification and layer thinning of MoS$_2$ in silver ionic (Ag$^+$) solution (Figure 1a). The experimental results show that under UV light exposure in Ag$^+$ solution environment, nanostructured silver ultrathin films (~2.9 nm) are uniformly decorated on 1L MoS$_2$ surfaces, meanwhile the thickness of MoS$_2$ from few-layer to bulk crystals could be thinned down to the atomic monolayer limit. In addition, the as-deposited Ag nanostructures can effectively control the *p*-type doping in 1L MoS$_2$ without causing any damage, as evidenced by *in-situ* electrical transport and *un-situ* Raman and photoluminescence (PL) measurements. Moreover, both silver nanostructure surface modification and layer thinning process are observed to start from the edges of MoS$_2$, revealing a unique UV-light-assisted defect-induced surface modification and thinning mechanism. Therefore, our work offers a simple and new technique for selectively controlling the doping and the layer thickness in TMDs, which can be extended to other van der Waals layered materials.

**RESULTS AND DISCUSSION**

We prepare MoS$_2$ flakes with different layer thicknesses by mechanical exfoliation and then transfer them onto the SiO$_2$ (285 nm)/Si substrates by dry transfer technique. Figure 1b show the optical image of a 1L MoS$_2$ flake, with its layer thickness confirmed by Raman frequency difference between $E_{2g}^1$ (384.7 cm$^{-1}$) and $A_{1g}$ (404.7 cm$^{-1}$) (Figure 1c). We find that 1L MoS$_2$ sample in Figure 1b has two edges with an angle of about 30°: one edge is long and straight (green arrows), while the other is relatively short (purple arrows). According to the previous studies,[29] the long and short edges correspond to the zigzag (ZZ) and armchair (AC) orientations of MoS$_2$, respectively (Figure 1b inset). Next, we immerse the sample in Ag$^+$ solution and perform UV light irradiation treatment (wavelength: 254 nm) (Figure 1a). After exposure with UV light for a fixed time (Figure 1d), the optical contrast change occurs from two edges of 1L MoS$_2$ (blue arrows), which can be more clearly identified in the inset of Figure 1d. The similar phenomenon has been observed in more 1L MoS$_2$ samples (Supporting Information), indicating that the appearance of optical contrast change is not limited to the types of MoS$_2$ edges.

There are two possible reasons for this optical contrast change in 1L MoS$_2$: surface corrosion or surface modification. To understand this phenomenon, we first perform Raman spectroscopy analysis. The Raman spectra of 1L MoS$_2$ taken at edge area (with optical change) and center area (without optical change) are similar (Figure 1e and Figures S2), consistent with that before UV irradiation (Figure 1c). In addition, there is no formation of any defect-related Raman peaks and other possible chemical bonds, such as Ag$_2$S (~244 cm$^{-1}$) and Mo-O (~820 cm$^{-1}$) (Supporting Information),[30-31] strongly confirming that the lattice structure of 1L MoS$_2$ at edge area (with optical change) is not damaged. Hence, the possibility of surface corrosion in 1L MoS$_2$ can be ruled out. Energy dispersive X-ray (EDX) analyses (Supporting Information) reveal that the change in optical contrast is due to the surface modification with Ag coatings on 1L MoS$_2$, as further evidenced by atomic force microscopy (AFM) studies. Figure 1f shows the AFM image for the blue dotted square in the inset of Figure 1d, which includes three regions: 1L MoS$_2$, Ag/1L MoS$_2$, and SiO$_2$/Si substrate. It is obvious that Ag/1L MoS$_2$ region is composed of a discontinuous nanostructured Ag thin film and a 1L MoS$_2$ underneath. In contrast, no Ag nanostructures are deposited on the SiO$_2$/Si substrate. The height difference between the crack in the area of Ag/1L MoS$_2$ and the substrate is measured to be about 0.7 nm (Figure 1f insert), which is in good agreement with the thickness of 1L MoS$_2$. Hence, the film thickness of as-deposited Ag nanostructures can be estimated, that is about 2.9 nm.

According to the observation in Figure 1, surface modification with Ag nanostructures prefers to initiate from the edges of 1L MoS$_2$. As we know, there are more defects at the edge than at the center area.[32] Previous studies have shown that the formation of defects within the 1L MoS$_2$ could introduce active sites, thereby leading to a significant improvement of the catalytic properties.[33] Therefore, in our case, we reasonably propose a UV light-assisted defect-induced surface modification mechanism in 1L MoS$_2$. To confirm this speculation, we prepare a 1L MoS$_2$ flake without edges by photolithography technique and then perform UV light treatment in Ag$^+$ solution (Supporting Information). As expected, there is no any Ag nanostructure deposition and no any change in the lattice structure for this 1L MoS$_2$ sample.

Next, we quantitively investigate the nanostructured Ag thin film deposition process on 1L MoS$_2$. Figure 2a shows different steps of surface modification under UV irradiation, where Ag nanostructures simultaneously nucleate from two edges of MoS$_2$ (green arrows, Stage 1), then expand and go inward (blue arrows, Stage 2), and finally reach the full coverage (Stage 3). We also have calculated the coverage of Ag nanostructures on the surface of MoS$_2$ using equation $(S_{Ag}/S_{MoS_2}) \times 100\%$, where $S_{Ag}$ and $S_{MoS_2}$ are the total areas of Ag nanostructures and pristine MoS$_2$, respectively. The fitting curve in Figure 2a demonstrates a typical parabola relationship between $S_{Ag}/S_{MoS_2}$ and UV irradiation time, similar to the photocatalytic behavior of Ag nanostructure deposition on other 2D materials.[34]

To gain a better understanding of the surface modification process and its effect on the electrical properties of MoS$_2$, we design a FET device in two-terminal configuration (Figure 2b) and perform *in-situ* electrical measurement. Figure 2c show the optical image of a 1L MoS$_2$ device on SiO$_2$/Si substrate fabricated by transferring an exfoliated sample on top of two parallel Au electrodes. The channel width and length are about 2 and 3 μm, respectively. Figure 2d compares the drain current ($I_d$) vs. drain-source voltage ($V_{ds}$) characteristics for 1L MoS$_2$ in Figure 2c after UV irradiation in Ag$^+$ solution for different time. The conductivity (σ) extracted from Figure 2d gradually decreases with increase in the UV irradiation time (or $S_{Ag}/S_{MoS_2}$ ratio) (Figure 2e), revealing a controllable carrier (electron or hole) doping in 1L MoS$_2$ by varying the Ag nanostructure modification. To confirm the doping effect, we have performed *in-situ* electrical transport measurement (Figure 2f). It can be seen that the pristine 1L MoS$_2$ FET exhibits a typical *n*-type semiconducting behavior. After modification with Ag nanostructures, $I_d$ (or σ) is significantly decreased in the whole gating range, representing a heavy hole (or *p*-type) doping. In addition, Ag nanostructure modification induces threshold voltage ($V_{th}$) shift toward the forward gate voltage ($V_{bg}$) direction (Figure 2e insert). This is another key indicator of *p*-type doping effect in 1L MoS$_2$.[35]

Optical characterization is also an effective approach to probe the doping effect in 1L TMDCs.[36] Figure 3a-3b analyze the room temperature PL spectra of 1L MoS$_2$ without and with Ag nanostructure deposition. We quantitatively fit the PL spectra using Gaussian model with three peaks, associated with B exciton peak at 2.02 eV, neutral exciton peak (A$^o$) at 1.90 eV, and negative trion peak (A$^-$) at 1.86 eV. The intensity ratios between A$^-$ and A$^o$ are 12.2 for 1L MoS$_2$ and 6.6 for Ag/1L MoS$_2$, suggesting a reduced electron doping in Ag/1L MoS$_2$. This optical behavior can be explained that the presence of Ag nanostructures results in the internal electron loss in 1L MoS$_2$. In addition, Raman spectra of 1L MoS$_2$ (Figure 1e) show that $A_{1g}$ peak position blue shifts by ~1 cm$^{-1}$ after deposition of Ag nanostructures, which further confirms the *p*-type doping effect.[37]

According to the previous reports,[31, 38] UV-ozone treatment is capable to induce the structural defects in 1L MoS$_2$ and then influence its optical and electrical properties (e.g., reduction of σ). To exclude the possibility of this occurrence in our case, we have investigated the structural and optoelectronic properties of 1L MoS$_2$ by UV light treatment in the air atmosphere (Figure 3c). Figure 3d shows the UV irradiation time (*t*) dependent Raman spectra for a 1L MoS$_2$ flake. As *t* increases, both the peak intensity and width of $E_{2g}^1$ (and $A_{1g}$) mode remain stable. Meanwhile, no significant signal change is observed in the time dependent PL spectra (Supporting Information), thereby suggesting no structural defects generated in 1L MoS$_2$. We then characterize the photo-response behavior of 1L MoS$_2$ FET using UV light as the excitation source. As shown in Figure 3e, when illustrated under 254 nm UV light, $I_d$ (or σ) is significantly increased by over one order of the magnitude throughout the gating range, which could return back to its original state when UV light is turned off. In addition, the current signal keeps stable even under UV light illustration for a long time (≥ 70 min) (Figure 3f). These experimental results provide strong evidence of neither structural degradation nor property change in 1L MoS$_2$ via UV irradiation.

Hence, the modulation of electrical properties of 1L MoS$_2$ in Figure 2 is mainly attributed to the surface modification of Ag nanostructures. Overall, UV light-assisted surface modification could effectively control the $S_{Ag}/S_{MoS_2}$ ratio and then the *p*-type doping level in 1L MoS$_2$.

How about UV light interaction with few-layer MoS$_2$ in Ag$^+$ solution environment? To answer this question, we mechanically exfoliate a bilayer (2L) MoS$_2$ flake (together with a 1L flake) on SiO$_2$/Si substrate (Figure 4a) and its layer thickness has been determined by Raman spectroscopy. After exposure with UV light in Ag$^+$ solution for a short time (< 5 min), the optical contrast at most areas of 2L MoS$_2$ has changed in brightness (Figure 4b), which is close to that of pristine 1L MoS$_2$. To understand this optical behavior, we have performed Raman spectroscopy analysis (Figure 4c-4e). It is found that the Raman frequency difference between $E_{2g}^1$ and $A_{1g}$ modes is changed from 22.4 cm$^{-1}$ (pristine area, Figure 4c) to 21.7 cm$^{-1}$ (reacting area, Figure 4d), strongly confirming that a layer thinning of MoS$_2$ from 2L to 1L occurs.[11]

To investigate the effect of UV irradiation time, we fabricate a 2L MoS$_2$ device and perform *in-situ* optical and electrical measurements (Figure 4f-4i). Figure 4f shows the optical images of 2L MoS$_2$ device after UV irradiation in Ag$^+$ solution for different time. In the initial stage (0 < $t$ ≤ 5 min), the optical contrast of the whole 2L channel becomes bright (similar to Figure 4b), which is attributed to the layer thinning effect. When $t$ > 5 min, 2L channel exhibits a similar optical behavior as that observed in 1L MoS$_2$ (Figure 2a), where the optical contrast becomes brighter from two edges of MoS$_2$ ($t$ ~ 10 min), then goes inward ($t$ ~ 15 min), and finally realizes the full color change ($t$ ≥ 20 min). Raman analyses in Figure 4e reveal that 1L MoS$_2$ formed in the initial stage remains stable even with a long time ($t$ = 20 min) UV irradiation in Ag$^+$ solution. Hence, we suggest that the surface modification with Ag nanostructures is the likely reason for the observed optical contrast change in the later time period. Figure 5g shows $S_{Ag}/S_{MoS_2}$ vs. $t$ curve for 2L MoS$_2$ device, which exhibits a parabola relation when $t$ > 5 min, in good agreement with that of 1L MoS$_2$ (Figure 2a). According to the above analyses, we conclude that selective layer thinning and surface modification could be realized in 2L MoS$_2$ (Figure 4f inset). Next, we carry out *in-situ* σ vs. *t* (Figure 4h) and electrical transport (Figure 4i) measurements. Both experimental results demonstrate that the electrical properties of 2L MoS$_2$ are well modulated via layer thinning ($0 < t \leq 5$ min) and surface modification ($5 < t \leq 20$ min), which could be explained by the thickness-dependent electron mobility[39] and the Ag deposition induced *p*-doping effect,[37] respectively.

To investigate the UV light interaction with multi-layer MoS$_2$ in Ag$^+$ solution, MoS$_2$ flakes with different thicknesses are exfoliated on SiO$_2$/Si substrate. A MoS$_2$ flake consisted of 1L to 6L, and their respective layer numbers, are shown in Figure 5a. The numbers of layers have been precisely verified by combing Raman spectroscopy (Figure 5b)[11] and optical contrast (black curve, Figure 5c).[40] After exposure with UV light in Ag$^+$ solution for a relatively long time (~ 60 min, Figure 5d), MoS$_2$ flake with different thicknesses all still exist. More interestingly, the optical contrast of MoS$_2$ ranging from 2L to 6L has changed into that is comparable to 1L region (red curve, Figure 5c), which indicates that laser thinning could also be realized in few- to multi-layer samples. Finally, we have extended our investigation to the thick-layer (TL) ones. Figure 5e compares the optical images of a TL MoS$_2$ flake after UV irradiation in Ag$^+$ solution for different time. As expected, layer thinning occurs in the TL sample. Similar to the surface modification process as in 1L MoS$_2$, laser thinning also starts from the edges of MoS$_2$ (Figure 5e inset). It is noted that the layer thinning rate in TL MoS$_2$ is significantly slower than that in few-layer ones, which can be utilized to fabricate thickness-dependent MoS$_2$ heterojunction devices.[41]

Based on the above analyses, we propose a physical model for better understating the layer thinning and surface modification in MoS$_2$ via UV irradiation in Ag$^+$ solution environment. For simplicity, here we take a 3L MoS$_2$ as an example (Figure 5f). First, the layer thinning (or etching) starts from the edges of top layer due to its high photocatalytic activity (Stage 1). When the top layer disappears, the second layer is exposed to the Ag$^+$ solution environment and begins to be etched in the same way (Stage 2). This process is repeated until the layered MoS$_2$ becomes into monolayer. Finally, Ag nanostructures are photo-deposited on the surface of 1L MoS$_2$ following the way as in Figure 2a (Stage 3). We believe that our developed UV light-assisted selective surface modification and layer thinning approach could be applicable to many other van der Waals layered materials.

**CONCLUSIONS**

In summary, we report the selective control of the surface modification and layer thickness of MoS$_2$ via UV light irradiation in an Ag$^+$ solution environment. By UV light treatment in Ag$^+$ solution, nanostructured silver ultrathin (~2.9 nm) films could be uniformly photo-deposited on 1L MoS$_2$, which can induce the controllable *p*-type doping without causing any lattice structural damage. On the other hand, few- to thick-layer MoS$_2$ could be efficiently thinned down to the monolayer limit with high crystallinity. Based on the *in-situ* electrical transport and *un-situ* optical spectroscopy analyses, we have proposed a UV light-assisted defect-induced surface modification and thinning mechanism. More importantly, the surface modification and layer thinning of MoS$_2$ can be selectively realized via the proposed single-platform approach, which is simple, convenient, and contactless. Our study thus presents a new strategy for tunning the intrinsic properties of TMDs, which is important for developing 2D material-based electronic and optoelectronic device applications.

**METHODS**

**Sample preparation.** The MoS$_2$ flakes with different layer thicknesses are mechanically exfoliated from their bulk materials (Six Carbon Technology, China) and then transferred onto the SiO$_2$ (285 nm)/Si substrates. Their layer thicknesses are identified by combining optical microscope (ECLIPSE LV150N, Nikon, Japan), Raman spectroscopy (Renishaw inVia), and AFM (MFP-3D, Oxford) measurements. The MoS$_2$ FET devices are fabricated by the standard photolithography process. Briefly, Au/Ti (10 nm/2 nm) electrodes are pre-patterned on the SiO$_2$/Si substrate, and then MoS$_2$ flakes on Gel-film are transferred on top of the pre-patterned Au/Ti electrodes using a dry transfer technique.[11]

**UV irradiation treatment of MoS$_2$ in Ag$^+$ solution.** The MoS$_2$ samples and FET devices are immersed into 3 mM AgNO$_3$ solution, and then exposed to UV light (wavelength: 254 nm, power density: 18 μW/cm$^2$) for a fixed irradiation time (Figure 1a). *In-situ* optical and electrical measurements are performed during the UV light irradiation process. For comparison, we have carried out UV light treatment and UV photo-response experiments in the air atmosphere.

**Characterizations.** The morphologies of MoS$_2$ and Ag/MoS$_2$ composites are characterized by optical microscope and AFM system. The structure and composition of our samples are analyzed by field emission scanning electron microscopy (JSM-7900F, Jeol, Japan) with EDX detection capability. The optical properties of MoS$_2$ samples are investigated by micro-Raman and PL system. Both Raman and PL signals are collected by focusing a 532 nm laser (power: 1.85 mW) onto the sample surface through a 50× objective. The electrical and photo-response measurements are carried out using semiconductor device analyzer (B1500A, Keysight, USA) in combination with a probe station (CM-4, Shenzhen Cindbest Technology Co., Ltd, China).

## ACKNOWLEDGEMENTS

This work was supported by the National Natural Science Foundation of China (NSFC) (Grant No. 12274051), and the Fundamental Research Funds for the Central Universities (Grant Nos. DUT21RC(3)032, DUT22ZK109). L.P. acknowledges the support from NSFC (Grant Nos. 51972039, 52272288). H.H. acknowledges the support from the Applied Basic Research Project of Liaoning Province, China (Grant No. 182022JH2/101300259), and the Dalian Science and Technology Innovation Fund, China (Grant No. 2022JJ12GX011). X.H. acknowledges the support from NSFC (Grant No. 62105053), and the Fundamental Research Funds for the Central Universities (Grant No. DUT21LK07).

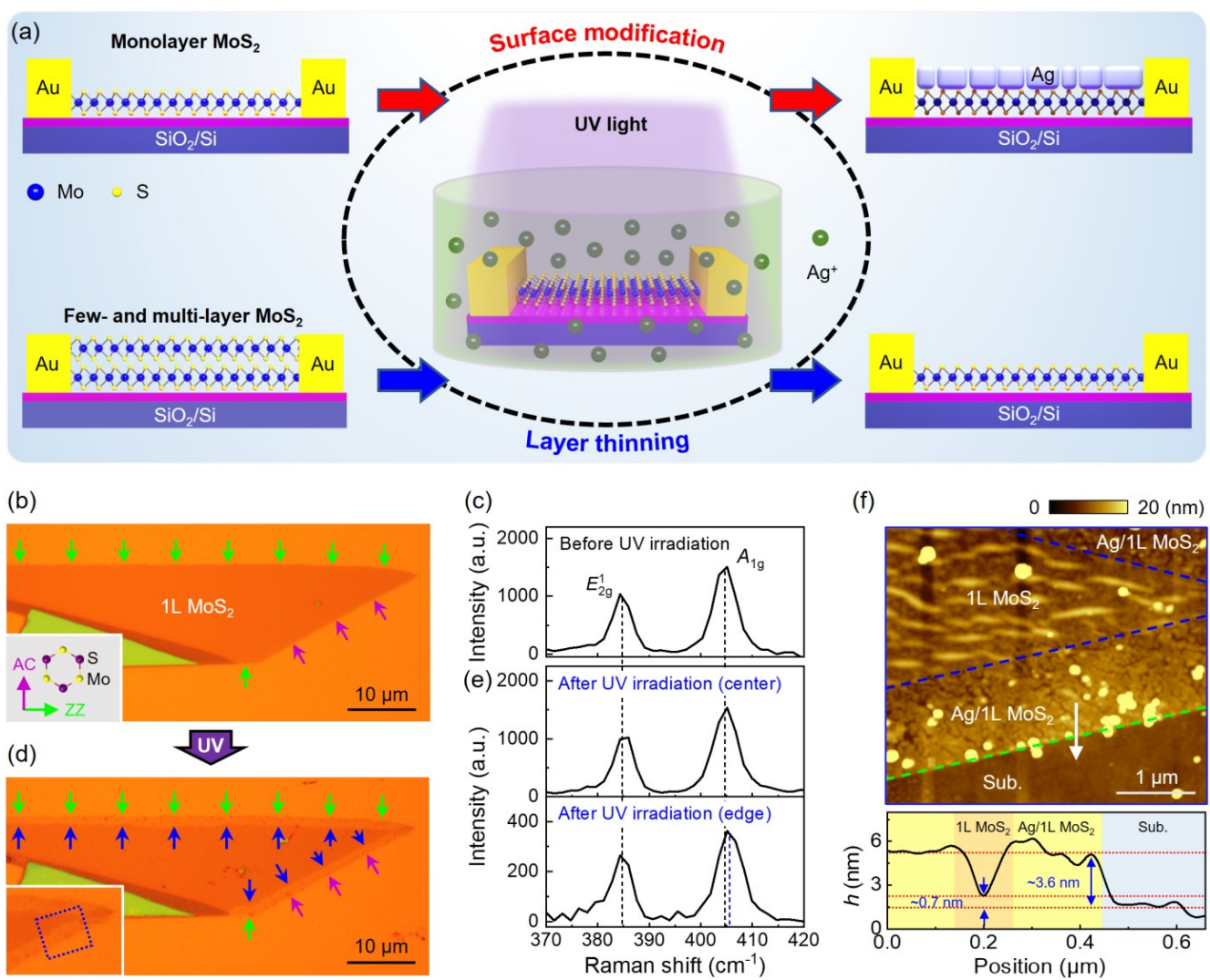

**Figure 1. Surface modification in 1L MoS₂.** (a) Schematic for selective surface modification and layer thinning of MoS$_2$ via UV irradiation in Ag$^+$ solution. (b and d) Optical images of a 1L MoS$_2$ flake (b) before and (d) after UV irradiation in Ag$^+$ solution for 20 min. Insets: The crystal orientation of MoS$_2$ (b) and enlarged optical image of the modified area (d). The green and purple arrows point to the edges of 1L MoS$_2$; the blue arrows point to the edges of modified area with Ag nanostructures. (c and e) Raman spectra of 1L MoS$_2$ sample shown in (b) and (d). (f) AFM image of the blue dotted square region in the inset of (d), which includes 1L MoS$_2$, Ag/1L MoS$_2$, and SiO$_2$/Si substrate. Lower inset: The height profile along the white solid line.

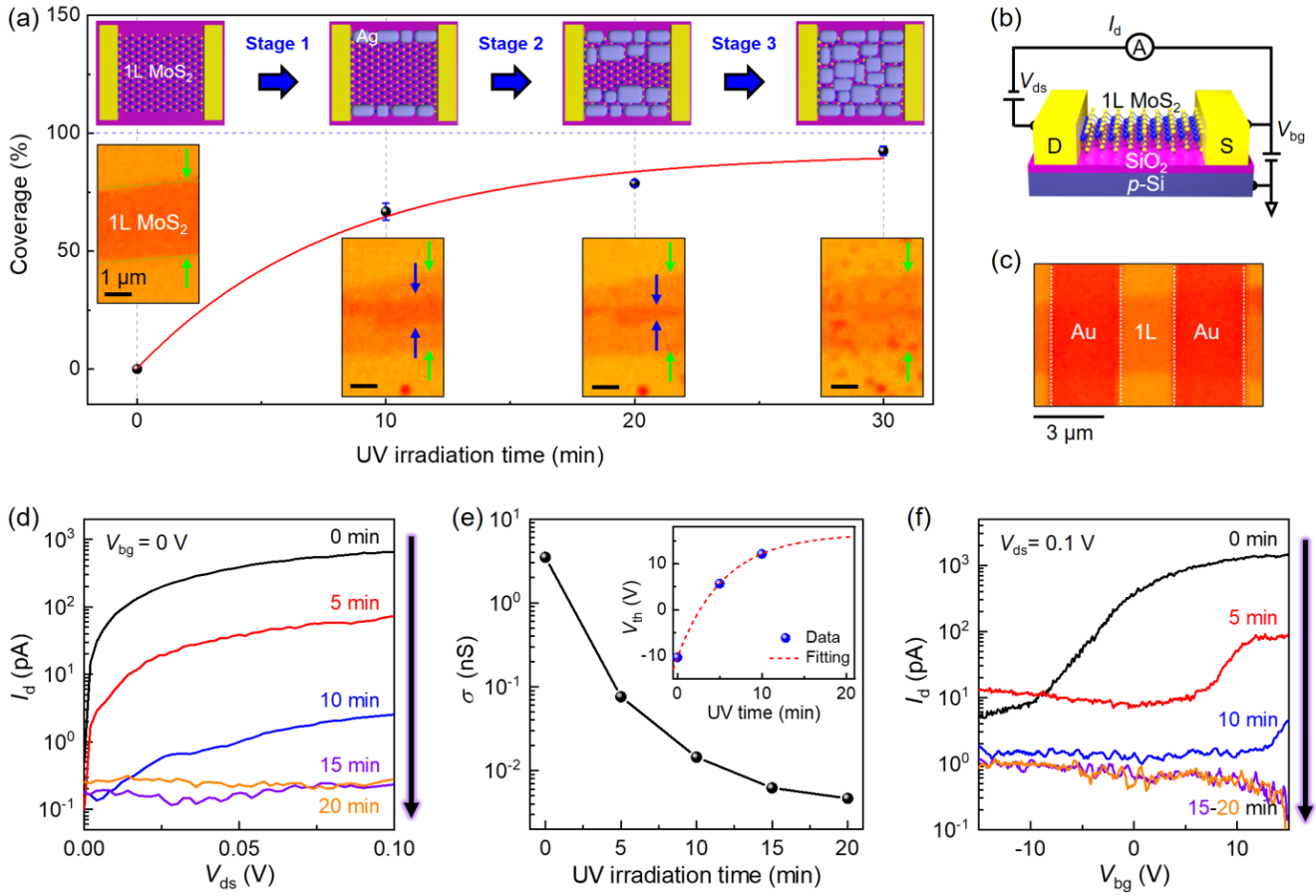

**Figure 2.** *In-situ* optical and electrical monitoring of the surface modification process in 1L MoS$_2$. (a) The coverage of as-deposited Ag nanostructures on 1L MoS$_2$ as a function of UV irradiation time. The dots and solid line are the experimental data and fitting curve, respectively. Lower insets: The corresponding optical images. The green and blue arrows point to the edges of pristine and reacting 1L MoS$_2$, respectively. Upper insets: Scheme of surface modification process in 1L MoS$_2$. (b) 1L MoS$_2$ FET device schematic. (c) Optical image of a 1L MoS$_2$ device, where Au electrodes are indicated by dotted lines. (d) *In-situ* $I_d$ vs. $V_{ds}$ ($V_{bg}$ = 0 V) characteristics of 1L MoS$_2$ device after UV irradiation in Ag$^+$ solution for different time. (e) The conductivity σ vs. UV irradiation time. Inset: The threshold $V_{th}$ extracted from (f) vs. UV irradiation time. (f) *In-situ* $I_d$ vs. $V_{bg}$ ($V_{ds}$ = 0.1 V) characteristics of 1L MoS$_2$ device after UV irradiation in Ag$^+$ solution for different time.

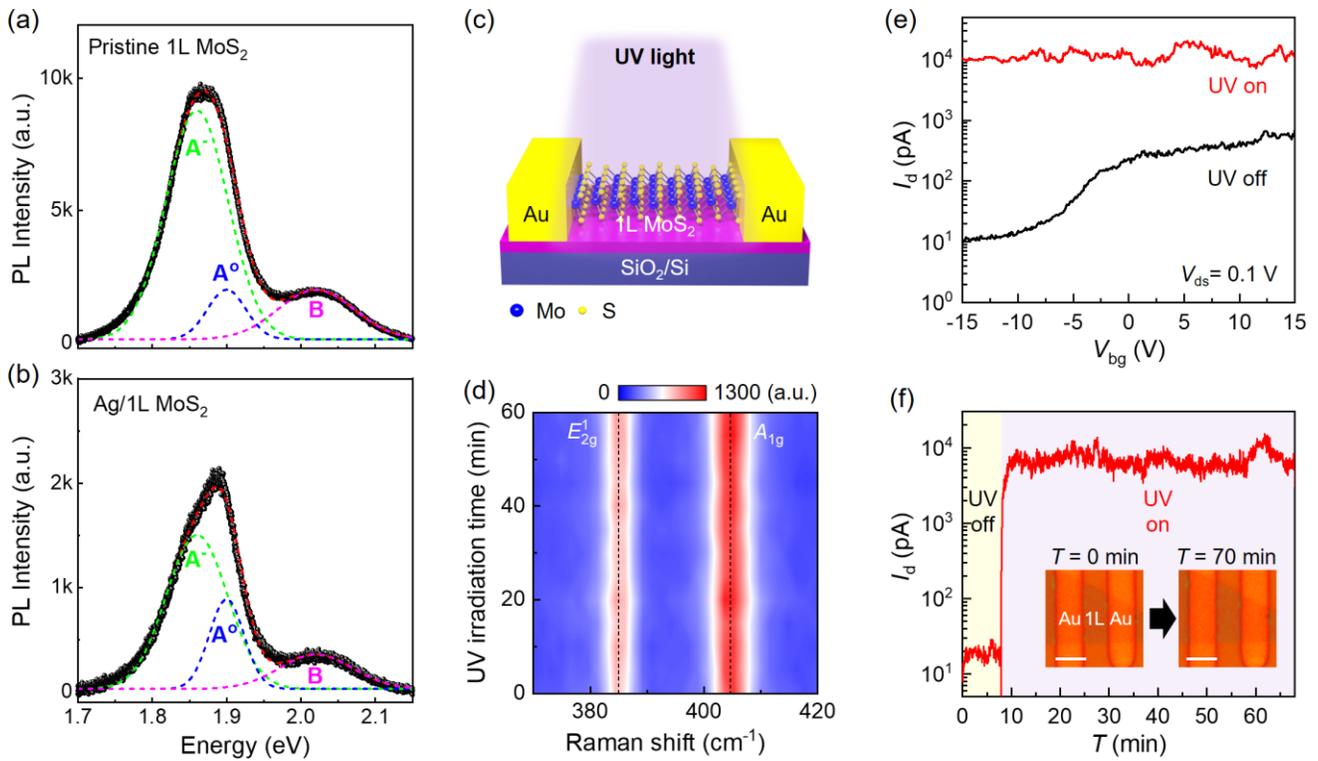

**Figure 3. The mechanism for tunning the electrical and optical properties of 1L MoS$_2$ via UV irradiation in Ag$^+$ solution.** (a and b) PL spectra for (a) pristine 1L MoS$_2$ and (b) Ag/1L MoS$_2$. The PL spectra are well fitted with three excitonic peaks: negative trion (A$^-$), neutral exciton (A$^o$), and B exciton. (c) Schematic of 1L MoS$_2$ device by UV light treatment in air. (d) Raman spectra of a 1L MoS$_2$ sample after UV irradiation for different time in air. (e) $I_d$ vs. $V_{bg}$ ($V_{ds}$ = 0.1 V) characteristics of a 1L MoS$_2$ device taken in dark (black curve) and under UV light exposure (red curve). (f) $I_d$ vs. time ($T$) relation ($V_{ds}$ = 0.1 V) for a 1L MoS$_2$ device (inset). Scale bar: 5μm.

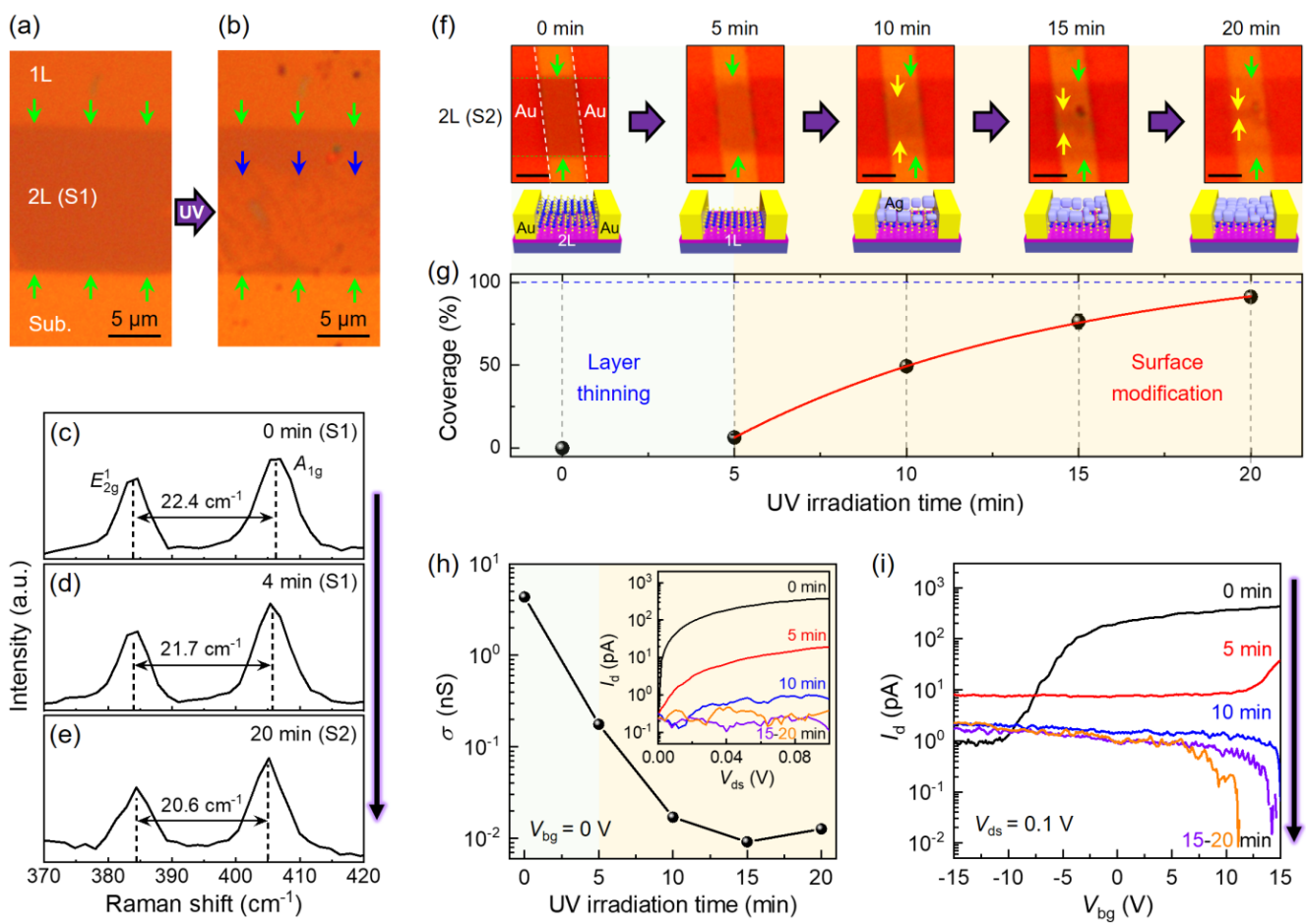

**Figure 4. UV light interaction with 2L MoS$_2$ in Ag$^+$ solution.** (a and b) Optical images of a 2L MoS$_2$ (S1) flake (a) before and (b) after UV irradiation in Ag$^+$ solution for 4 min. The green and blue arrows point to the edges of pristine and reacting 2L MoS$_2$, respectively. (c-e) Raman spectra for 2L MoS$_2$ (c) before and after UV irradiation in Ag$^+$ solution for (d) 4 min and (e) 20 min. (f-i) *In-situ* optical and electrical monitoring of UV light interaction with 2L MoS$_2$ (S2) device in Ag$^+$ solution. (f) Optical images of S2 device after UV irradiation in Ag$^+$ solution for different time. The green and yellow arrows point to the edges of pristine 2L MoS$_2$ and the edges of Ag/MoS$_2$ region, respectively. Scale bar: 2 μm. Lower insets: The corresponding device schematics. (g) The coverage ($S_{Ag}/S_{MoS_2}$) vs. UV irradiation time relation. The dots and solid line are the experimental data and fitting curve, respectively. (h) σ *vs.* $t$ and $I_d$ *vs.* $V_{ds}$ (inset) characteristics. (i) $I_d$ *vs.* $V_{bg}$ ($V_{ds}$ = 0.1 V) characteristics of S2 device after UV irradiation in Ag$^+$ solution for different time.

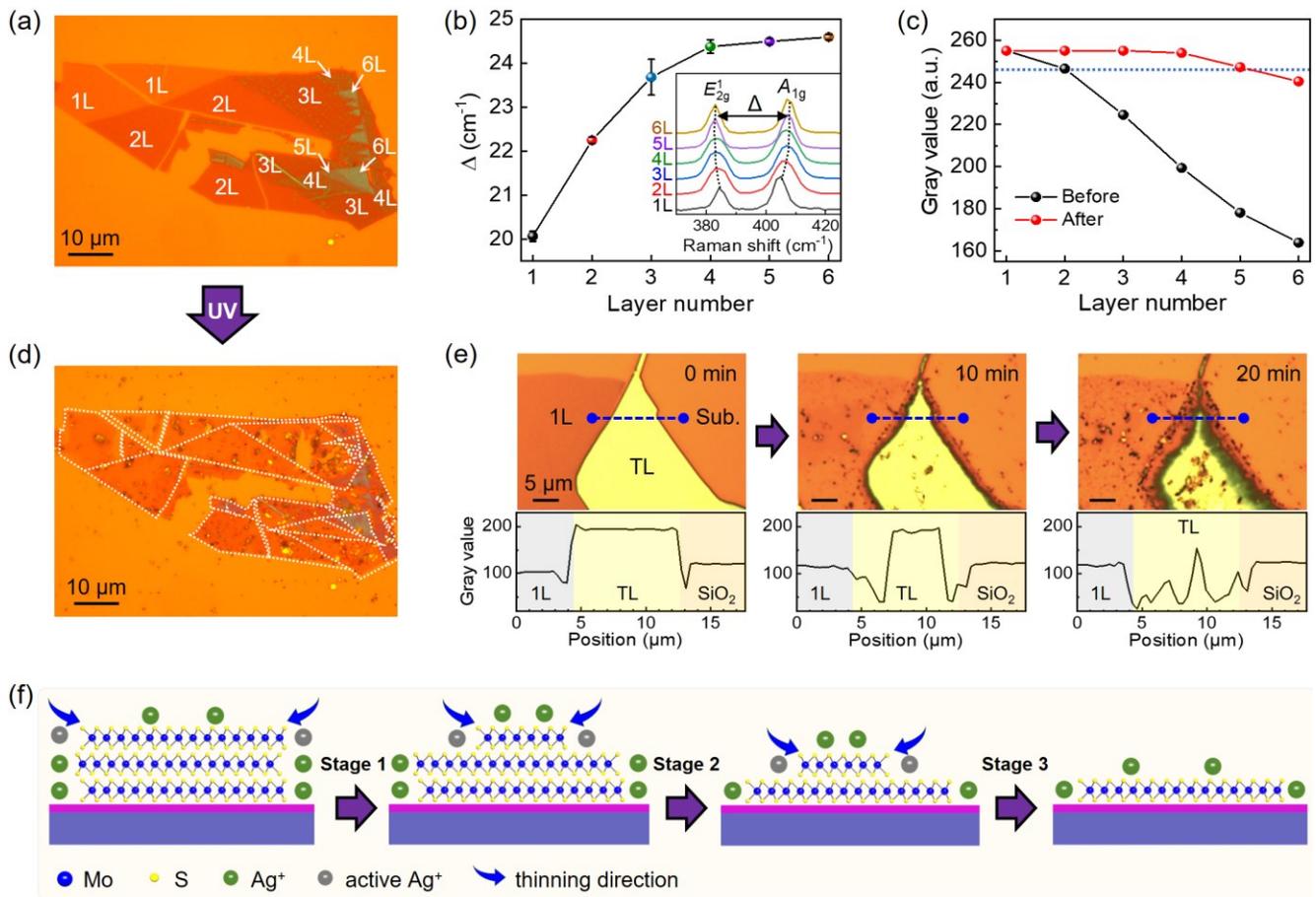

**Figure 5. UV light interaction with multi- to thick-layer MoS$_2$ in Ag$^+$ solution.** (a and d) Optical images of a MoS$_2$ flake consisted of 1L to 6L (a) before and (d) after UV irradiation in Ag$^+$ solution for 60 min. (b) Raman frequency difference (Δ) as a function of MoS$_2$ layer thickness. Inset: Raman spectra for the MoS$_2$ in (a) with different layer thicknesses. (c) The gray value extracted from the sample in (a and d) as a function of layer thickness. The blue dotted line represents 2L MoS$_2$. (e) Optical images of a thick-layer MoS$_2$ under UV irradiation for different time. Lower insets: The corresponding gray-value profiles along the blue dashed lines. (f) The proposed mechanism of layer thinning in MoS$_2$ via UV irradiation in Ag$^+$ solution environment.